\documentclass[conference]{IEEEtran}
\IEEEoverridecommandlockouts
\usepackage{cite}
\usepackage{amsmath,amssymb,amsfonts}
\usepackage{algorithmic}
\usepackage{graphicx}
\usepackage{textcomp}
\usepackage{xcolor}

\usepackage{comment}
\usepackage{seqsplit} 
\usepackage{xcolor} 
\usepackage{tcolorbox}
\usepackage{enumitem} 
\usepackage{verbatim}
\usepackage{varwidth}
\usepackage{listings}
\usepackage{booktabs}  
\usepackage{caption}   
\usepackage{tabularx}  
\usepackage{amsmath} 
\usepackage{colortbl}     
\usepackage{csquotes} 
\usepackage{upquote} 
\usepackage{hyperref} 
\usepackage{lineno} 

\usepackage{blindtext}

\newcommand{\toolname}{SynergyRCA} 
\newcommand{\stategraph}{StateGraph}
\newcommand{\metagraph}{MetaGraph}
\newcommand{\gpt}{GPT-4o}

\newcommand{\rcl}{\emph{Triage}} 
\newcommand{\rclplain}{Triage}

\newcommand{\cqg}{\emph{PathQueryGen}} 
\newcommand{\cqgplain}{PathQueryGen} 

\newcommand{\ds}{\emph{StateChecker}} 

\newcommand{\rcg}{\emph{ReportGen}} 
\newcommand{\rcgplain}{ReportGen}

\newcommand{\ies}{\emph{ReportQualityChecker}} 
\newcommand{\iesplain}{ReportQualityChecker} 

\newcommand{\kbs}{Kubernetes} 

\newcommand{\chaimengfei}{foobar2} 
\newcommand{\baishen}{foobar3} 
\newcommand{\yanghao}{foobar4} 
\newcommand{\dushiting}{foobar5} 

\begin{document}

\title{Simplifying Root Cause Analysis in Kubernetes with StateGraph and LLM\\
}


\author{\IEEEauthorblockN{Yong Xiang}
\IEEEauthorblockA{
\textit{Tsinghua University}\\
Beijing, China \\
xiangyong13@mails.tsinghua.edu.cn}
\and
\IEEEauthorblockN{Charley Peter Chen}
\IEEEauthorblockA{
\textit{Harmonic Inc}\\
Vancouver, Canada \\
charley.chen@harmonicinc.com}
\and
\IEEEauthorblockN{Liyi Zeng}
\IEEEauthorblockA{
\textit{Peng Cheng Laboratory}\\
ShenZhen, China \\
zengly@pcl.ac.cn}
\and
\IEEEauthorblockN{Wei Yin}
\IEEEauthorblockA{
\textit{Tsinghua University}\\
Beijing, China \\
yw@mail.tsinghua.edu.cn}
\and
\IEEEauthorblockN{Xin Liu}
\IEEEauthorblockA{
\textit{Tsinghua University}\\
Beijing, China \\
liuxin19@mails.tsinghua.edu.cn}
\and
\IEEEauthorblockN{Hu Li}
\IEEEauthorblockA{
\textit{Unaffiliated}\\
Beijing, China \\
lihu723@gmail.com}
\and
\IEEEauthorblockN{Wei Xu}
\IEEEauthorblockA{
\textit{Tsinghua University}\\
Beijing, China \\
weixu@tsinghua.edu.cn}
}

\maketitle

\begin{abstract}
\kbs{}, a notably complex and distributed system, utilizes an array of controllers to uphold cluster management logic through state reconciliation. Nevertheless, maintaining state consistency presents significant challenges due to unexpected failures, network disruptions, and asynchronous issues, especially within dynamic cloud environments. These challenges result in operational disruptions and economic losses, underscoring the necessity for robust root cause analysis (RCA) to enhance \kbs{} reliability.
The development of large language models (LLMs) presents a promising direction for RCA. However, existing methodologies encounter several obstacles, including the diverse and evolving nature of \kbs{} incidents, the intricate context of incidents, and the polymorphic nature of these incidents. 
In this paper, we introduce \toolname{}, an innovative tool that leverages LLMs with retrieval augmentation from graph databases and enhancement with expert prompts.
\toolname{} constructs a \stategraph{} to capture spatial and temporal relationships and utilizes a \metagraph{} to outline entity connections.
Upon the occurrence of an incident, an LLM predicts the most pertinent resource, and \toolname{} queries the \metagraph{} and \stategraph{} to deliver context-specific insights for RCA. We evaluate \toolname{} using datasets from two production \kbs{} clusters, highlighting its capacity to identify numerous root causes, including novel ones, with high efficiency and precision. \toolname{} demonstrates the ability to identify root causes in an average time of about two minutes and achieves an impressive precision of approximately 0.90.

\end{abstract}

\begin{IEEEkeywords}
Root Cause Analysis, Large Language Models, Incident Management, Retrieval-Augmented Generation, Graph, GPT-4o
\end{IEEEkeywords}

\section{Introduction}
Cluster orchestration platforms such as \kbs{}~\cite{k8s}, Borg~\cite{verma2015large}, ECS~\cite{melissaris2022elastic}, and Twine~\cite{tang2020twine} are widely utilized for managing an array of systems across diverse domains and applications~\cite{cncf23}. \kbs{}, in particular, is a complex and distributed system made up of numerous components spread across multiple layers. 
It operates through a set of \textit{controllers}~\cite{controllers_and_reconciliation} that implement cluster management logic based on the \textit{state reconciliation principle}~\cite{controllers_and_reconciliation, burns2016borg}. These controllers continuously monitor the cluster state, striving to align the \textit{current state} with the \textit{desired state}.
However, maintaining consistency across these states is challenging due to unexpected failures~\cite{gu2023acto, sun2024anvil}, network interruptions~\cite{sun2022automatic}, and asynchrony issues~\cite{diouf2020byzantine}, particularly when operating within complex, dynamic, and distributed cloud environments. Numerous documented failure incidents have resulted in significant operational disruptions and economic losses~\cite{k8sfail, cebula10weird, madhu2022preventing, zookeeperop, casskop, tidbop}.

Root cause analysis (RCA) is crucial for the swift recovery of service availability and the enhancement of \kbs{} reliability. 
The rapid development and successful application~\cite{jain2022jigsaw, mastropaolo2022using, fan2022improving, mastropaolo2021studying} of LLMs present a promising avenue for RCA improvement. 
Ahamed et al.~\cite{ahmed2023recommending} conducted the first large-scale empirical study evaluating LLMs' effectiveness in RCA, 
while Zhang et al.~\cite{zhang2024automated}  demonstrated that using in-context examples can improve root cause recommendation quality. 
Additionally, Chen et al.~\cite{chen2024automatic} suggested employing in-context examples with a retrieval-augmented generation (RAG) framework for root cause categorization.

However, applying AI-based RCA in \kbs{} introduces several challenges. 
Firstly, the diverse array of \kbs{} incidents, varying in versions and configurations, continually evolves, complicating the RCA process. The fine-tuning method described by Ahamed et al.~\cite{ahmed2023recommending} is costly and challenging to update continuously, risking hallucinations~\cite{chen2024automatic}. Rule-based AI-enhancements like K8sGPT~\cite{k8sgpt}, which encodes Site Reliability Engineer (SRE) experience into analyzers, are difficult to maintain and require substantial expert labor for version updates.

Secondly, interpreting the wide range of incident contexts further complicates RCA. Chen et al.~\cite{chen2024automatic} utilized RAG to supply LLMs with incident-related data specific to a private service, and Zhang et al.~\cite{zhang2024automated} found that 20-shot in-context learning could achieve best performance. In our research, we bypass the \kbs{} incident handler and avoid directly importing examples into prompts, which restricts the LLM's performance with limited examples. Instead, we propose structuring domain-specific runtime data in a graph database for dynamic querying to enhance LLM effectiveness.

Thirdly, the polymorphic nature of incident message semantics poses an additional challenge for LLM analysis. For instance, in the error message \textit{ ``Error: cannot find volume `foobar1' to mount into container `foobar2',''} the volume could be a \texttt{ConfigMap}, \texttt{Secret} or \texttt{PVC}. Drishti et al.~\cite{goel2024x} demonstrated that augmenting the LLM prompt with data from various stages of the software development lifecycle (SDLC) can improve RCA. Building upon this approach, we propose incorporating unstructured knowledge into the LLM through expert prompts.

In this paper, we introduce \emph{\toolname{}}, an innovative tool that utilizes off-the-shelf  LLMs enhanced with retrieval augmentation from a graph database that captures the runtime states of \kbs{} clusters. This tool is further refined with expert prompts. \toolname{} is widely applicable across various \kbs{} cluster instances and capable of analyzing a wide range of failure messages within incidents. We automatically construct a \emph{\stategraph{}}  to capture spatial and temporal relationships of entities within a \kbs{} instance and employ a \emph{\metagraph{}} to map the interconnections among various entity kinds. These graphs not only facilitate data management for LLM input but also act as a protective barrier, preventing direct LLM operations on the production cluster.

Upon the occurrence of a new incident, an LLM module, improved with expert prompts, predicts the most pertinent resource for examination. \toolname{} then queries the \metagraph{} to uncover dependencies among resource entities and formulates a graph query to identify state dependencies on related entities within the \stategraph{}. This information is fed into the LLM, which subsequently generates a root cause report and suggests remediation commands.

\toolname{} employs the essential principle of state reconciliation to identify root causes by comparing states through three key steps: (1) verifying the existence of the current state, such as checking for a directory's presence on an NFS server; (2) ensuring the correctness of these states, such as confirming the resource quota is not exhausted; and (3) identifying discrepancies among states, such as verifying if a \texttt{Pod} and \texttt{PVC} exhibit a matching ``Bound" status.

We implement \toolname{} using \gpt{} and the Neo4j graph database~\cite{neo4j}. We assess its performance with datasets from two production \kbs{} clusters of different versions, covering periods of one week and six months, respectively. \toolname{} successfully identifies 18 and 20 types of root causes in these datasets, notably discovering five new types in the second dataset, underscoring its effectiveness. \toolname{} operates efficiently, pinpointing root causes within an average of two minutes, and accurately identifies the root cause in 90\% of cases.

This paper presents a comprehensive approach to LLM-based RCA with three key contributions:
\begin{itemize}

\item \emph{Model RCA with Graph-based RAG}: We are pioneers in integrating graph databases with LLMs to generate RCA reports, advancing the accuracy and contextual understanding in RCA significantly.

\item \emph{End-to-End LLM-based RCA System}: We introduce \toolname{}, an innovative end-to-end solution that automatically builds graphs and leverages LLMs to precisely identify root causes and generate insightful RCA reports.

\item \emph{Real-World Evaluation on Kubernetes Clusters}: We demonstrate the practical utility and dependability of \toolname{} in two real-world Kubernetes cluster scenarios.

\end{itemize}

The rest of the paper is organized as follows: Section~\ref{related} reviews related work, Section~\ref{sec:design} details the design of \toolname{}, Section~\ref{evaluation} discusses evaluation experiments and results, Section~\ref{discuss} addresses limitations and potential future work, and finally, Section~\ref{conclude} concludes the paper.

\section{Related Work}
\label{related}
\subsection{Incident Management with AI}
Traditional research in incident lifecycle automation has extensively employed machine learning (ML) techniques to enhance processes such as 
triaging~\cite{azad2022picking, chen2019empirical, chen2019continuous}, 
incident linking~\cite{chen2020identifying, ghosh2024dependency}, 
diagnosis~\cite{bansal2020decaf, luo2014correlating, nair2015learning},
and mitigation~\cite{jiang2020mitigate}.
Regarding RCA tasks, some studies have concentrated on identifying the problem's root cause~\cite{lin2016log, yuan2019approach, lee2023eadro, zhang2023robust}, while others have focused on determining the components involved~\cite{ikram2022root, wang2023root, li2022causal, li2021practical}. 
Marco et al.~\cite{barletta2024mutiny} investigated real-world \kbs{} failures and developed a framework for executing fault injection campaigns to replicate these patterns. 
CausalIoT~\cite{wang2023iot} introduced an anomaly detection system using device interaction graphs to identify abnormal states in IoT environments. 
RAPMiner~\cite{liu2022rapminer} provided an anomaly localization mechanism for CDN systems, employing classification power-based redundant attribute deletion to eliminate non-root cause attributes. 
Despite their utility, these methods often involve intricate feature extraction processes and lack cross-platform applicability and task flexibility~\cite{zhang2024survey}.
Some approaches, such as topology graph-based analyses~\cite{brandon2020graph, wu2020microrca, wu2020performance}, reconstruct graphs to model application topology and pinpoint potential anomaly root causes. Whereas these methods primarily focus on system components, our approach with \stategraph{} offers finer granularity by concentrating on entities and their interrelations.

\subsection{LLMs in Software System Reliability}
Large Language Models (LLMs) are revolutionizing the landscape of software systems by offering innovative solutions for various tasks, including
code generation~\cite{chen2021evaluating, xu2022systematic}, 
code repair~\cite{fan2022improving, joshi2023repair}, 
vulnerability repair~\cite{fu2022vulrepair},
and bug fixing~\cite{mastropaolo2021studying}.
Our research focuses on enhancing software system reliability by using LLMs to automate and improve the accuracy of the RCA process.

Ahamed et al.~\cite{ahmed2023recommending} conducted the first large-scale study assessing the effectiveness of LLMs for RCA and mitigation of production incidents. 
Zhang et al.~\cite{zhang2024automated} introduced an in-context learning approach for automated RCA and performed an extensive analysis of over 100,000 incidents to tackle the high costs associated with fine-tuning LLMs. 
Drishti et al.~\cite{goel2024x} demonstrated that incorporating contextual data from various stages of the software development lifecycle (SDLC) enhances performance on critical tasks. 
Our work synthesizes these methodologies by employing a prompt-based approach with graph-based RAG and expert enhancements, thereby avoiding the high costs of extensive fine-tuning and numerous in-context learning demonstrations.

RCAgent~\cite{wang2023rcagent} presented a tool-augmented LLM autonomous agent designed for secure industrial use by deploying models internally to prioritize security over performance. 
K8sGPT~\cite{k8sgpt} integrated SRE experience into its analyzers and utilized LLMs exclusively for generating natural language outputs, without providing a complete end-to-end solution. 
As a result, these approaches do not fully harness the capabilities of external models like GPT-4o, as achieved in our work.

LogConfigLocalizer~\cite{shan2024face} introduced a two-stage LLM-based strategy to identify root-cause configuration properties from Hadoop logs.
RCACopilot~\cite{chen2024automatic} combined RAG and in-context learning to predict root cause categories, furnishing explanatory narratives by consolidating diagnostic information from incident handlers.
Roy et al.~\cite{roy2024exploring} evaluated a ReAct agent for the dynamic collection of supplementary incident-related diagnostic data.
While these methods emphasize failure localization, classification, and data collection, our approach utilizes RAG with graphs to conduct comprehensive, end-to-end RCA and offer remediation suggestions. 
Our work notably introduces the \stategraph{}, enabling LLMs to utilize structured runtime data for better contextual understanding in RCA.
Conversely, GraphRAG~\cite{graphrag} created knowledge graphs from raw text for community-oriented summaries but lacked the capability to capture rapidly changing system states, rendering it unsuitable for our \stategraph{}.

\section{\toolname{} Design}
\label{sec:design}

We introduce \toolname{}, our LLM-based approach for RCA, which employs retrieval-augmented generation (RAG) from graph to enhance LLM performance. Section~\ref{subsec:design_overview} provides an overview of our methodology. Section~\ref{subsec:llm_rca_design} details the integration of knowledge into LLMs for effective RCA. Finally, Section~\ref{subsec:stategraph_design} and Section~\ref{subsec:metagraph_design} explore the organization of knowledge within the \stategraph{} and \metagraph{}, respectively.

\subsection{Design Overview}
\label{subsec:design_overview}

\subsubsection{Core Concepts}
This section briefly introduces the fundamental concepts related to graphs utilized in \toolname{}. 
Within a \kbs{} cluster and its related ecosystems, an \emph{Entity} refers to a distinct and identifiable object in the system. A \emph{Snapshot} captures the state of an entity at regular intervals, offering insights into system dynamics. If the entity is not an \texttt{Event} resource, the snapshot is referred to as the \emph{State}, with uppercase convention applied to denote Snapshots.
We construct a \emph{\stategraph{}}  to capture the dependencies among entities across both temporal and spatial dimensions. In the \stategraph{}, an \emph{entity vertex} is created for each entity, and a \emph{snapshot vertex}  is defined for each snapshot, specifically a \emph{STATE vertex} for State. A \emph{statepath} represents a path through the \stategraph{} that includes only entity vertices, excluding any snapshot vertices. To provide a concise overview of the relationships among various \emph{kinds} of entities, we derive a \emph{\metagraph{}}  from the \stategraph{}, depicting the overall structural connections. A \emph{metapath} within the \metagraph{} comprises paths typically consisting only of entity kinds, excluding snapshot kinds.
These graphs are stored in Neo4j~\cite{neo4j}, a graph database well-suited for managing relationships. LLMs are employed to generate Cypher queries, Neo4j’s intuitive query language~\cite{cypher}, to facilitate data access and manipulation.

\subsubsection{\toolname{} Workflow }
\begin{figure}[tb]
  \centering
  \includegraphics[width=0.5\textwidth]{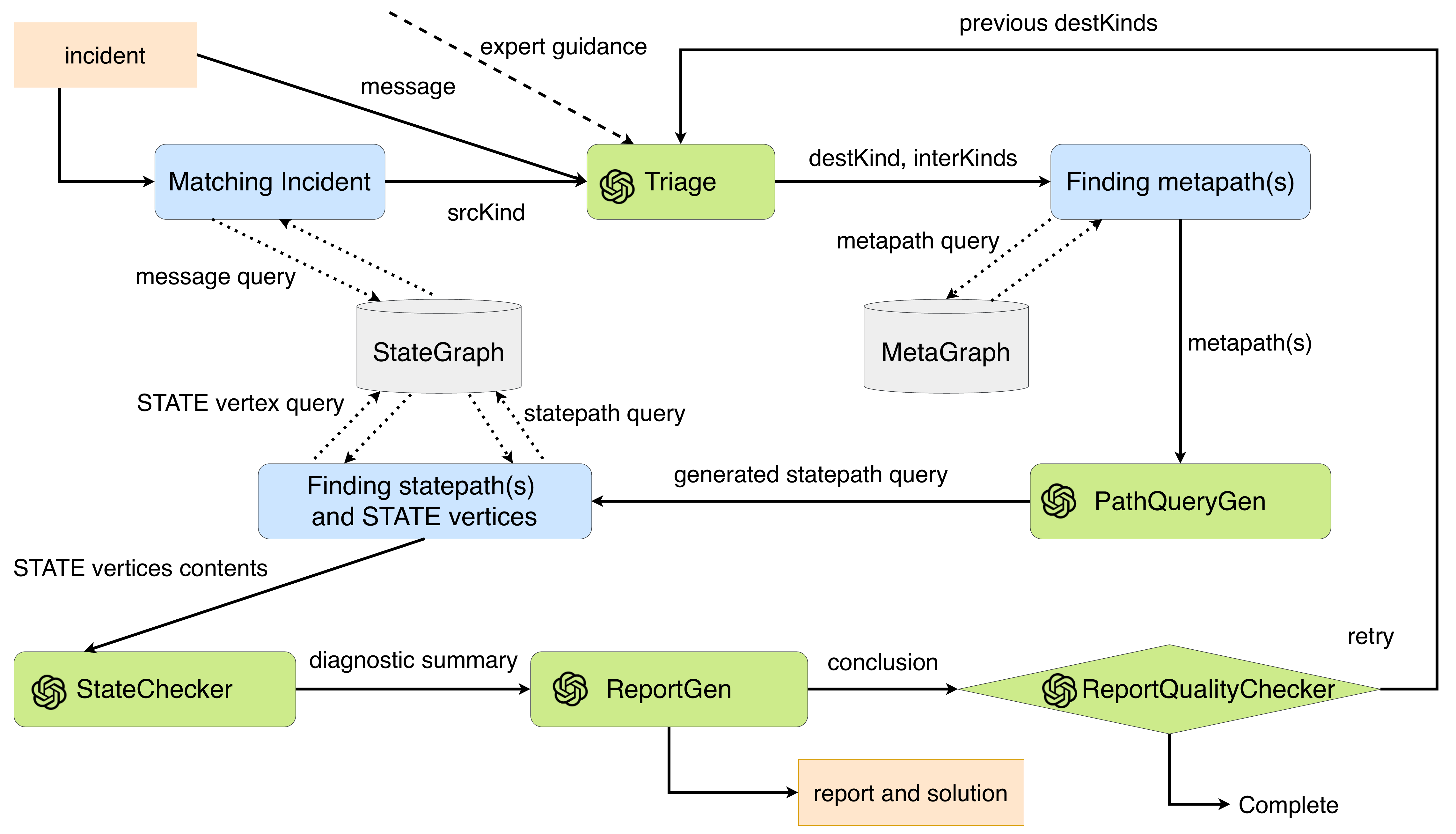}
  \caption{Overview of \toolname{}'s  Retrieval Augmented LLM Workflow} 
  \label{fig:synergyRCA_overview} 
\end{figure}

Fig. \ref{fig:synergyRCA_overview} illustrates the structure of our proposed methodology. Initially, we capture the spatial and temporal relationships among entities within the \stategraph{} and further delineate the dependencies between various entity kinds using the \metagraph{}. These graphs form the foundational knowledge base for subsequent queries that augment LLM prompts.

The analysis procedure is managed by a standard \emph{driver} program. Upon the occurrence of a new incident, the driver queries the \stategraph{} to match the incident’s message, namespace, and timestamp, thereby identifying the source entity kind (the \emph{srcKind}, e.g., \texttt{Job} or \texttt{ReplicaSet}) involved in the incident. This srcKind, along with the incident message, is fed into the \rcl{} module, which utilizes an LLM to predict the root cause entity kind (the \emph{destKind}, e.g., \texttt{ResourceQuota})  and any relevant resource kinds  (the \emph{interKinds}, e.g., \texttt{Pod} and \texttt{Namespace}).

Subsequently, the driver queries the \metagraph{} to locate one or more metapaths connecting the srcKind to the destKind, potentially traversing through interKinds. Each metapath is transformed into a Cypher query by the \cqg{} module, with the aid of an LLM for query generation. Executing these queries against the \stategraph{} yields a statepath (e.g., \texttt{Job1-Namespace1-ResourceQuota1}).

For each entity along the statepath, the driver re-queries the \stategraph{}  to retrieve the current state of the entity (e.g., the details of \texttt{RESOURCEQUOTA1}). The state information for each entity is then input into the \ds{} module, which employs an LLM to extract a diagnostic summary relevant to the incident message.

Based on these diagnostic summaries, the \rcg{} module, powered by an LLM, generates a root cause report and suggests remediation commands. Furthermore, the \ies{} module utilizes an LLM to assess how effectively the report’s conclusion explains the error message and determines if further investigation is required.

If the current conclusion does not adequately explain the error message, the incident is re-routed back to the \rcl{} module to identify a new destKind and interKinds for further analysis. This iterative process continues until a satisfactory explanation and remediation plan are achieved, or until the maximum number of trials is reached.

\subsection{LLM-based RCA Design}
\label{subsec:llm_rca_design}

In this section, we outline how to effectively integrate graph and expert knowledge into LLMs through prompts to conduct RCA. Following the workflow depicted in Fig.~\ref{fig:synergyRCA_overview}, we describe the design of each LLM module.

\subsubsection{\textbf{\rcl{}}}
Upon the occurrence of an incident, predicting the potential root cause location is essential. The \rcl{}  prompt, illustrated in Fig.~\ref{fig:RootCauseLocatorPrompt}, serves this purpose. 
Firstly, the LLM is provided with a specific starting point (srcKind) by querying the  \stategraph{}  to match the incident (line 5). 
Subsequently, the LLM is tasked with identifying all relevant resource kinds (interKinds) that might contribute to the incident (lines 6--7). 
Thirdly, the LLM determines a destination kind (destKind) most likely to be the root cause (lines 8--16). 
Finally, an in-context learning approach~\cite{zhang2024automated} instructs the LLM to strictly follow a JSON output format (lines 17--41).

\begin{figure}[tb]
    \centering
    \begin{tcolorbox}[width=0.5\textwidth, colframe=black, colback=white, left=4mm, right=2mm, top=2mm, bottom=2mm]
     \scriptsize
     
\setlength\linenumbersep{2 mm} 
\setcounter{linenumber}{1}
\begin{internallinenumbers}

\textbf{\#\#\# Task Description}   

Refer to the predefined resource kinds and naming convention. Analyze the following Kubernetes error message that includes \{involved\_object\}. 
 
 \textbf{\#\#\# Analysis Steps}   
     

1. Recognize the \{involved\_object\} as the starting point of the issue.

2. Identify the most critical Kubernetes native and external resources relevant to the problem from the predefined resource kinds.

3. Determine the `destKind', which is the resource kind most directly related to the problem. 
Guidelines:

(1) Provide only one `destKind'; do not list multiple kinds.

(2) The `destKind' should be the most crucial one for the error message. For example, If a quota is exceeded,  the `destKind' should be `ResourceQuota' .

(3) The `destKind' must be in the relevant resources list and predefined resource kinds.


(4) If `destKind' is an external kind, it should always be in lowercase.

(5) If multiple possible kinds, infer the best match using naming conventions.

\textbf{\#\#\# Output Formatting}

Output the findings in JSON format encapsulated within triple backticks and the `json' specifier. The JSON output should not contain additional descriptions and must follow the given structure:
 
\begin{verbatim}
```json
{"SourceKind": "{involved_object}",
 "DestinationKind": "destKind", 
 "RelevantResources": ["Resource1", "Resource2", ...,  
	                      "{involved_object}", "destKind"]
}
```
\end{verbatim}
  
\textbf{\#\#\# Example}


Sample Error Message:

\begin{verbatim}
```
Error creating: pods "es-cronjob-1607245800-kmtgd" is 
forbidden: exceeded quota: ...   
```
\end{verbatim}

Sample Output:

\begin{verbatim}
```json
{"SourceKind": "Job",
 "DestinationKind": "ResourceQuota",
 "RelevantResources": ["Job", "Pod", "Namespace", 
	                      "ResourceQuota"],
}
```
\end{verbatim}

\textbf{\#\#\# Input}

Analyze the following error message ensuring `destKind' and `RelevantResources' are strictly limited to the provided lists.  \{error\_message\}

 \end{internallinenumbers}  

      \end{tcolorbox}
    \caption{Prompt snippet in \rcl{}}
    \label{fig:RootCauseLocatorPrompt}
\end{figure}



To make \rcl{} more effective, we incorporate entity taxonomy, an aspect of graph knowledge, along with expert guidance into its prompts, as shown in Fig.~\ref{fig:graph_expert_knowledge}.
Initially, a predefined list of existing entity kinds is given to the LLM to prevent the creation of non-existent kinds through hallucination (lines 2--5). 
Additionally, we provide expert guidance to prioritize the prediction focus on specific destKinds that may be the ``usual suspects" in a specific cluster.
For example, we direct attention to specific external kinds like \texttt{nfs}, \texttt{container}, and \texttt{image} (lines 7--11). 
Finally, we apply naming conventions to help the LLM infer destKind, and users can customize these conventions for their specific \kbs{} clusters (lines 12--16).
For instance, in handling the error message \textit{``Error: cannot find volume `gen-white-list-conf' to mount into container `es-crontab-job'.''}  the presence of `-conf' in the name indicates it is more likely a \texttt{ConfigMap} than a \texttt{Secret} or \texttt{PVC}.

\begin{figure}[tb]
    \centering
    \begin{tcolorbox}[width=0.5\textwidth, colframe=black, colback=white, left=4mm, right=2mm, top=2mm, bottom=2mm]
    \scriptsize

\setlength\linenumbersep{2 mm} 
\setcounter{linenumber}{1}
\begin{internallinenumbers}

\textbf{\#\#\# Graph Knowledge of Entity Taxonomy}

The predefined Kubernetes native resource kinds and external resource kinds are listed below:

- k8s native resource kinds: \{nativeKinds\}

- External resource kinds: \{externalKinds\}

\textbf{\#\#\# Expert Knowledge}

Please note:

- For the k8s native kinds, focus on the commonly used kinds: [`ConfigMap', `CronJob', `DaemonSet', `Deployment', `Endpoints', `Image', `Job', ...].

- For the external kinds, prioritize focusing on: [`nfs', `container', `image', `hostpath'].

- Naming Conventions: 

(1) k8s external kinds are always lowercase, while most k8s native kinds are capital case.

(2) `-conf' is more likely a ConfigMap, while `-cert' and `-token' are more likely a Secret.

 \end{internallinenumbers}      
   
      \end{tcolorbox}
    \caption{Prompt snippet to import graph and expert knowledge}
    \label{fig:graph_expert_knowledge}
\end{figure}


\rcl{} is a critical module that leverages LLM capabilities to minimize the need for user experience in \kbs{}. As a result, \rcl{} can accurately identify \texttt{nfs} as the destKind and recommend \texttt{PVC} and \texttt{PV} as potential interKinds for \texttt{FailedMount-NoSuchFileDir}  issues.
\rcl{} can be executed in a zero-shot manner using a sufficiently large LLM, such as \gpt{}, due to the widespread usage of \kbs{} and the availability of common error data from various \kbs{} versions. Additionally, deployment-specific errors are managed within the expert-prompt (lines 6--16).

\subsubsection{\textbf{\cqg{}}}

We utilize the code generation capabilities of LLMs to develop \cqg{}, which translates a metapath into a Cypher query. \cqg{} enhances the metapath by prepending the sequence ({``\texttt{EVENT} \textrightarrow \texttt{Event} \textrightarrow \texttt{srcKind}''}) to it. For example, the extended metapath for issues like \texttt{FailedMount-NoSuchFileDir} would be:

\begin{lstlisting}[basicstyle=\ttfamily\footnotesize]
HasEvent, Event, EVENT, metadata_uid;
ReferInternal, Event, Pod, involvedObject_uid;
ReferInternal, Pod, PersistentVolumeClaim, spec_volumes_persistentVolumeClaim_claimName;
ReferInternal, PersistentVolume, PersistentVolumeClaim, spec_claimRef_uid;
UseExternal, PersistentVolume, nfs, spec_nfs_path;
\end{lstlisting}

\cqg{} initiates by processing the error message and constructs a MATCH-WHERE clause for each edge, subsequently returning the nodes and relationships (akin to vertices and edges) along the path. For example, for the {``\texttt{PV} \textrightarrow \texttt{PVC}''} edge, it generates the following Cypher clause:

\begin{lstlisting}[basicstyle=\ttfamily\footnotesize]
MATCH (pv:PersistentVolume)-[r4:ReferInternal]->(pvc:PersistentVolumeClaim)
WHERE r4.key = 'spec_claimRef_uid'
\end{lstlisting}

\begin{figure}[tb]
    \centering
    \begin{tcolorbox}[width=0.5\textwidth, colframe=black, colback=white, left=4mm, right=2mm, top=2mm, bottom=2mm]
     \scriptsize
    
\setlength\linenumbersep{2 mm} 
\setcounter{linenumber}{1}
\begin{internallinenumbers}    
    
 \textbf{\#\#\# Analysis Steps}   
 
3. Chain MATCH Clauses Based on the Metapath:

        (1) Continue the query by adding MATCH clauses for each part of the provided metapath. For each segment of the metapath, use the node type (srcKind and destKind) as the label for the source and destination node. Use the relationship type (relType) as the label for the connecting relationship, and apply a WHERE clause based on the `key' property value (propertyValue) specified for that relationship:
\begin{verbatim}
MATCH (startNode:srcKind)-[r1:relType]->(node1:destKind)
WHERE r1.key = 'propertyValue'
\end{verbatim}
	
        (2) For consecutive relationships, increment the relationship alias sequentially to use unique identifiers such as r1, r2, r3, etc. This ensures clarity when multiple relationships are present in the MATCH pattern:
\begin{verbatim}
MATCH (node1:srcKind)-[r2:relType]->(node2:destKind)
WHERE r2.key = 'propertyValue'
\end{verbatim}
	... and so on for additional relationships.
        
        (3) Ensure to use the same node alias for each node type, particularly if that node type appears in multiple relationships to maintain consistency. For example:
\begin{verbatim}
MATCH (evt: EVENT),
MATCH (n1:Event)-[r1:HasEvent]->(evt: EVENT),
MATCH (n1:Event)-[r2:ReferInternal]->(n2: Pod)
\end{verbatim}
	...

\textbf{\#\#\# Example}

Here’s an example for clarity: 
...

 \end{internallinenumbers}

%
%
%
%
%
%
%
%
%

      \end{tcolorbox}
    \caption{Prompt snippet in \cqg{}}
    
    \label{fig:CypherQueryGeneratorPrompt}
\end{figure}

%
%
%


Fig.~\ref{fig:CypherQueryGeneratorPrompt} depicts the essential steps (lines 2--20) of \cqg{}, which also employs an in-context learning approach (lines 22--23). We focus on generating Cypher queries specifically for statepaths due to their flexibility, whereas other Cypher queries tend to be simple and static.

\cqg{} leverages the knowledge within the \metagraph{} by utilizing a metapath. Even though LLMs like \gpt{} possess extensive knowledge of \kbs{}, they can still make errors when connecting specific kinds. For instance, \gpt{} might erroneously suggest a  {``\texttt{PV} \textrightarrow \texttt{Node} \textrightarrow \texttt{nfs}''} path that does not exist. The metapath ensures that \cqg{} generates accurate and executable Cypher queries in the \stategraph{} with high confidence.

\subsubsection{\textbf{\ds{}}}
We developed a powerful module, \ds{},  to extract diagnostic summaries from complex entity \emph{States}, leveraging the semantic understanding capabilities of LLMs. Traditional methods often face challenges due to the intricate structure and data types present in States. This complexity is further complicated by varying keys across different entity kinds, and even within a single kind. \ds{} excels in extracting the most pertinent information, providing significant advantages in navigating these complexities.

\begin{figure}[tb]
  \centering
  \includegraphics[width=0.5\textwidth]{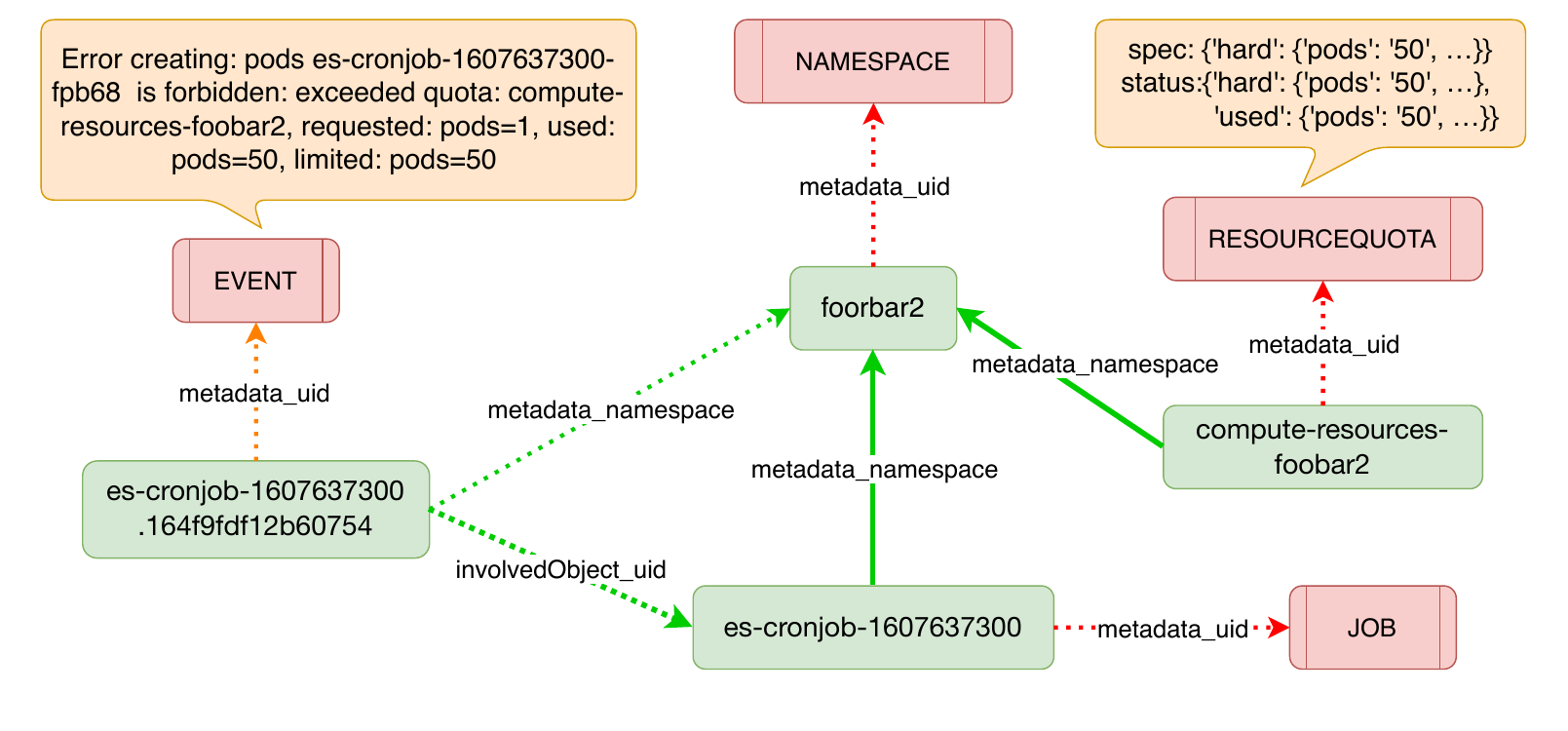}
 \caption{ An example statepath of \texttt{FailedCreate\-ExceedQuotaJob} with \texttt{Event} and \texttt{Snapshots} extended.}
  \label{fig:ExceedQuota-Job}
\end{figure}

We execute the generated Cypher query in \stategraph{} and get one or more instanced paths, known as \emph{statepaths}.
An example statepath of \texttt{FailedCreate-ExceedQuotaJob} ({\texttt{Job} \textrightarrow \texttt{Namespace} \textrightarrow \texttt{ResourceQuota}}, with \texttt{Event} and \texttt{Snapshots} extended) is illustrated in Fig.~\ref{fig:ExceedQuota-Job}.  
We then query each entity in the statepath to access its STATE vertex through a HasState edge (e.g., {\texttt{Job} \textrightarrow \texttt{JOB}}), with the content stored as a JSON string termed StateJSON.

We then provide the \ds{} with portions of the StateJSON, typically the ``spec" and ``status" fields, along with the error message, and instruct it to summarize key observations using the prompt snippet shown in Fig.~\ref{fig:DiagnosticSummarizerPrompt}.
\begin{figure}[tb]
    \centering
    \begin{tcolorbox}[width=0.5\textwidth, colframe=black, colback=white, left=4mm, right=2mm, top=2mm, bottom=2mm]
    \scriptsize

\setlength\linenumbersep{2 mm} 
\setcounter{linenumber}{1}
\begin{internallinenumbers}

\textbf{\#\#\# Analysis Steps}

   3. Conduct an evaluation to determine if there is any evidence (e.g., misconfigurations or errors) in the JSON fields, especially those which could align with the nature of the provided error message.
   
        - If the error message seems to relate to the JSON data, clarify the connection and identify any anomalies or errors in the data.
        
        - If the error message appears to be unrelated to the JSON data, clearly acknowledge this finding.
        
        - **Important Rule**: 
        
        (1) Your analysis must strictly adhere to the factual data provided in the JSON string. Do NOT create or fabricate any new JSON snippets. 
        
        (2) If the provided JSON cannot explain the root cause of the error message, clearly state that the JSON does not explain it. Do NOT infer the root cause from the error message, and do NOT attempt to justify or rationalize the error message.
        
        (3) Report any discrepancies between the error message and the JSON data. Avoid making incorrect observation based on mismatched or incorrect data.
        
        ...
        
\textbf{\#\#\# Output Formatting}

Summarize key observations and any issues discovered in the JSON data in a concise manner (limit within 200 words).

	Key points to include:

	(1) **Key Observations**: Highlight the most relevant findings ONLY based on the provided JSON data for the error message.   	


	...
	
   	(3) **Relevant JSON Fragments**: Include only the most relevant parts or fragments from the  JSON data that are crucial for understanding the issue. 
 	
	...
	    
        (b) Extract only the deepest fields directly related to the error message instead of including the full JSON structure.
	\begin{verbatim}
	```json
	{"spec":{"hard":{"pods":"40"}},
	 "status":{"hard":{"pods":"40"},"used":{"pods":"40"}}}
	```
	\end{verbatim}

 \end{internallinenumbers}      
                      
      \end{tcolorbox}
     \caption{Prompt snippet in \ds{}}
    \label{fig:DiagnosticSummarizerPrompt}
\end{figure}

The \ds{} assesses whether the error message correlates with the fields in the StateJSON. If they are related, it clarifies the connection and extracts the most pertinent JSON fragments; if they are not, it acknowledges their irrelevance (lines 2--16). It subsequently summarizes the key observations within 200 words (lines 18--33), as demonstrated in the diagnostic summary for \texttt{RESOURCEQUOTA}  in Fig.~\ref{fig:DiagnosticSummary-ResourceQuota}.

\begin{figure}[tb]
    \centering
    \begin{tcolorbox}[width=0.5\textwidth, colframe=black, colback=white, left=4mm, right=2mm, top=2mm, bottom=2mm]
     \scriptsize

\setlength\linenumbersep{2 mm} 
\setcounter{linenumber}{1}
\begin{internallinenumbers}

RESOURCEQUOTA(25f8e934-6e18-5b9c-bee4-b4231a9d01a6): 

Key Observations:

1. The error message indicates exceeding the pod quota in the `\baishen{}' namespace.

2. The JSON object is a ResourceQuota for the `\baishen{}' namespace.

3. The spec and status fields in the JSON confirm a limit of 50 pods and show that 50 pods are already in use.

Summary of Issues:

The error message is directly related to the JSON data. The ResourceQuota for the `\baishen{}' namespace specifies a hard limit of 50 pods, and the status shows that 50 pods are currently in use. Therefore, any attempt to create additional pods will exceed the quota and be forbidden.

Relevant JSON Fragments:

\begin{verbatim}
```json
{"spec":{"hard":{"pods":"50"}},
 "status":{"hard":{"pods":"50"},"used":{"pods":"50"}}}
```
\end{verbatim}

The JSON data clearly aligns with the error message by showing that the pod quota limit has been reached.

 \end{internallinenumbers}     
 
      \end{tcolorbox}
    \caption{An example diagnostic summary generated by \ds{}.}
    \label{fig:DiagnosticSummary-ResourceQuota}
\end{figure}



To minimize hallucinations, we design the prompt to require the LLM to strictly adhere to the factual data present in the StateJSON, avoiding any fabrication or inference, even if discrepancies exist (lines 9--16). This approach prevents the LLM from inferring root causes based solely on the error message. If no STATE vertex is available for an entity, we directly include a key observation noting this absence in the \ds{}, bypassing the previous prompt.

\subsubsection{\textbf{\rcg{}}}
Utilizing the diagnostic summary for each entity along the statepath, we employ three fundamental rules to ascertain the root cause of the error. These rules, rooted in the principle of state reconciliation, are applicable to systems beyond \kbs{}:

\begin{enumerate}
\item Each entity should have a current state represented by a STATE vertex.
\item The states of entities should be consistent (e.g., if a \texttt{Job} tries to create a \texttt{Pod}, the \texttt{ResourceQuota} should have remaining quota).
\item Any inconsistency due to data discrepancy should be clearly reported.
\end{enumerate}

We developed \rcg{}  to produce both a root cause report (lines 5--10) and recommend remediation commands (lines 11--15), as guided by the prompt in Fig.~\ref{fig:ReportCommandGeneratorPrompt}. A sample report is shown in Fig.~\ref{fig:report-command-exceedquota-job}, where the report accurately pinpoints the \texttt{ResourceQuota} as the root cause (lines 1--24) and suggests a specific kubectl command with the appropriate name and namespace (lines 25--28). This design builds on work by Ahmed et al.~\cite{ahmed2023recommending}, which demonstrated that GPT-3.x models enhance their ability to propose mitigation plans when the root cause is clearly identified.

\begin{figure}[tb]
    \centering
    \begin{tcolorbox}[width=0.5\textwidth, colframe=black, colback=white, left=4mm, right=2mm, top=2mm, bottom=2mm]
     \scriptsize

\setlength\linenumbersep{2 mm} 
\setcounter{linenumber}{1}
\begin{internallinenumbers}

\textbf{\#\#\# Task Description}

Based on the previous analysis of [\{kinds\}], summarize the root cause of the error message, and pinpoint out the most relevant parts.

\textbf{\#\#\# Analysis Steps}

    1. For each kind, faithfully summarize the findings based on evidences/facts only, and do not  include any suspicions that are not verified. Then provide a score (0 \textasciitilde 10/10) to indicate how relevant it is to the error message.

    2. Moreover, provide an overall score (0 \textasciitilde 10/10) to indicate how well the conclusion (and detailed summary if needed) can explain the root cause of the error message.

    3. Furthermore,  provide a resolution for the error with kubectl or bash command if applicable. Note: include crucial details such as resource names, IDs, and numbers that are pertinent to understanding the cause. The kubectl/bash command should incorporate the actual resource names, or namespaces, to achieve precision in execution.

    ...

 \end{internallinenumbers}  
                           
      \end{tcolorbox}
    \caption{Prompt snippet in \rcg{}}
    \label{fig:ReportCommandGeneratorPrompt}
\end{figure}

    

\begin{figure}[tb]
    \centering
    \begin{tcolorbox}[width=0.5\textwidth, colframe=black, colback=white, left=4mm, right=2mm, top=2mm, bottom=2mm]
    \scriptsize

\setlength\linenumbersep{2 mm} 
\setcounter{linenumber}{1}
\begin{internallinenumbers}

\begin{verbatim}
{"summary": [
  {"kind": "Job",
   "explanation": "The Job object specifies creating a 
       new pod in the 'foobar3' namespace, which is at 
       its pod quota limit.",
   "relevance_score": "9/10"
  },
  {"kind": "Namespace",
   "explanation": "The Namespace 'foobar3' is active but
       lacks details on resource quotas, providing no 
       direct evidence for the error.",
   "relevance_score": "2/10"
  },
  {"kind": "ResourceQuota",
   "explanation": "The ResourceQuota for the 'foobar3'
       namespace shows a hard limit of 50 pods, and the 
       'used' count is already at 50.",
   "relevance_score": "10/10"
  }],
 "conclusion": "The error message is caused by the 
       'foobar3' namespace reaching its pod quota limit 
       of 50, as confirmed by the ResourceQuota and Job
       objects.",
 "overall_score": "9/10",
 "resolution": "To resolve the error, increase the pod
       quota in the namespace 'foobar3': kubectl patch
       resourcequota compute-resources-foobar3 -n foobar3
       -p '{\"spec\":{\"hard\":{\"pods\":\"60\"}}}'",
}
\end{verbatim}

%
 
  \end{internallinenumbers}

      \end{tcolorbox}
    \caption{An example RCA report and recommended solution for \texttt{FailedCreate-ExceedQuotaJob}.}
    
    \label{fig:report-command-exceedquota-job}
\end{figure}




\subsubsection{\textbf{\ies{}}}

Finally, we assess the generated root cause report’s effectiveness in explaining the error message and determine the need for further investigation. Using the prompt in Fig.~\ref{fig:InvestigationEstimatorPrompt}, we direct the \ies{} to assign a score (lines 1–10) indicating whether additional investigation is necessary. This scoring approach, rather than a direct true/false decision, benefits from a longer chain-of-thought (CoT)~\cite{wei2022chain} for reasoning. 
If the driver operates interactively, user validation can serve as an alternative to the \ies{}.

\begin{figure}[tb]
    \centering
    \begin{tcolorbox}[width=0.5\textwidth, colframe=black, colback=white, left=4mm, right=2mm, top=2mm, bottom=2mm]
     \scriptsize

\setlength\linenumbersep{2 mm} 
\setcounter{linenumber}{1}
\begin{internallinenumbers}

\textbf{\#\#\# Analysis Steps}

Finally, determine if further investigation is needed.
Scoring and Investigation Criteria:
    
    1. If the conclusion alone can directly explain the root cause of the error message, score it above 9/10, and set ``further\_investigation": False.
    
    2. If the conclusion cannot solely explain the root cause, but in combination with a detailed summary, they together can explain the root cause, score it above 7/10 and set ``further\_investigation": False.
    
    3. If the conclusion and summary can only partially explain the root cause or are merely relevant to it, score it below 5/10 and set ``further\_investigation": True.
    
 \end{internallinenumbers}      

     \end{tcolorbox}
    \caption{Prompt snippet in \ies{}}
    \label{fig:InvestigationEstimatorPrompt}
\end{figure}
    

\subsection{Constructing \stategraph{} from Running \kbs{} Cluster}
\label{subsec:stategraph_design}

This section introduces \emph{\stategraph{}}, a graph structure designed to capture the spatial and temporal relationships among entities within a \kbs{} environment. \stategraph{} enables the identification of connections among entities at specific timestamps and the tracking of entity state evolution over time. We begin by defining the key concepts fundamental to the design of \stategraph{}  and then detail the construction process.

\subsubsection{Basic Concepts in \stategraph{}}  
  
\paragraph{Entities and Snapshots}
  
In the \kbs{} domain, an \emph{entity} represents a unique and identifiable object or concept within the system. Examples of entities include \texttt{Pods}, \texttt{PVCs}, and NFS directories. Entities are classified into two categories: \emph{k8s-native} entities, which are inherent to \kbs{} (e.g.,  \texttt{Pod}, \texttt{ReplicaSet}, \texttt{Job}, and \texttt{CronJob}) and stored in etcd, and \emph{k8s-external} entities, which are not native to \kbs{} but are still relevant (e.g., \texttt{containers}, \texttt{images}, and NFS directories).

Entities may consist of one or more fields. An entity with a single field (e.g., \texttt{IP}  address) is deemed \emph{atomic}; if it has multiple fields, it is considered \emph{composite}. Many entities, such as k8s-native entities like \texttt{Pods} or \texttt{PVCs} are composite and typically include fields like uid, name, namespace, and kind. Similarly, an NFS directory may have fields like path and server. Entities are uniquely identified by an identifier (e.g., uid) or a field combination (e.g., path and server).

We periodically collect data to capture the state of an entity at each timestamp, known as a \emph{Snapshot}. For example, querying etcd for Pods every 5 minutes provides snapshots of the Pod’s state at these intervals. However, rapid changes within an interval may result in inconsistent snapshots.

\paragraph{Vertices and Edges}

The \stategraph{} consists of two vertex types: \emph{entity vertices} and \emph{snapshot vertices}.
Lowercase letters represent k8s-external entities, capitalized letters denote k8s-native entities, and uppercase letters indicate snapshots.
For example, \texttt{POD} represents a snapshot of a \texttt{Pod}. If the snapshot is not for an \texttt{Event} resource, it is also referred to as a \emph{STATE vertex}. 

\begin{figure}[tb]
  \centering
  \includegraphics[width=0.5\textwidth]{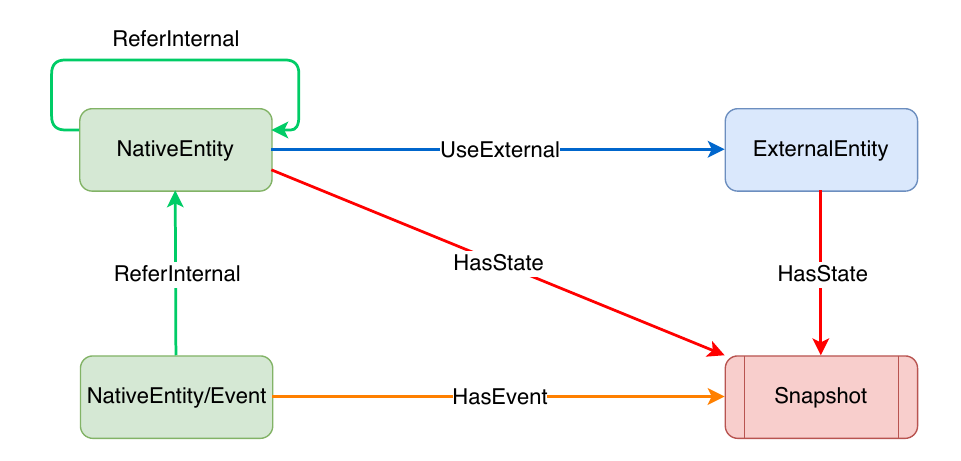}
  \caption{Schematic diagram of \stategraph{}}
  \label{fig:EntitySnapshotGraph}
\end{figure}

\stategraph{} features two types of Entity-Entity edges. Edges labeled \emph{ReferInternal} indicate native-to-native connections, while \emph{UseExternal} denotes native-to-external connections. Entity-Snapshot edges also have two types: due to  \texttt{Event} objects in etcd functioning like logs, we use the special \textit{HasEvent} type; all other connections, whether native or external, are labeled \textit{HasState}. These concepts and relationships form the foundation of the \stategraph{}, as illustrated in Fig.~\ref{fig:EntitySnapshotGraph}.

\subsubsection{Constructing the \stategraph{}}
\begin{figure}[tb]
  \centering
  \includegraphics[width=0.5\textwidth]{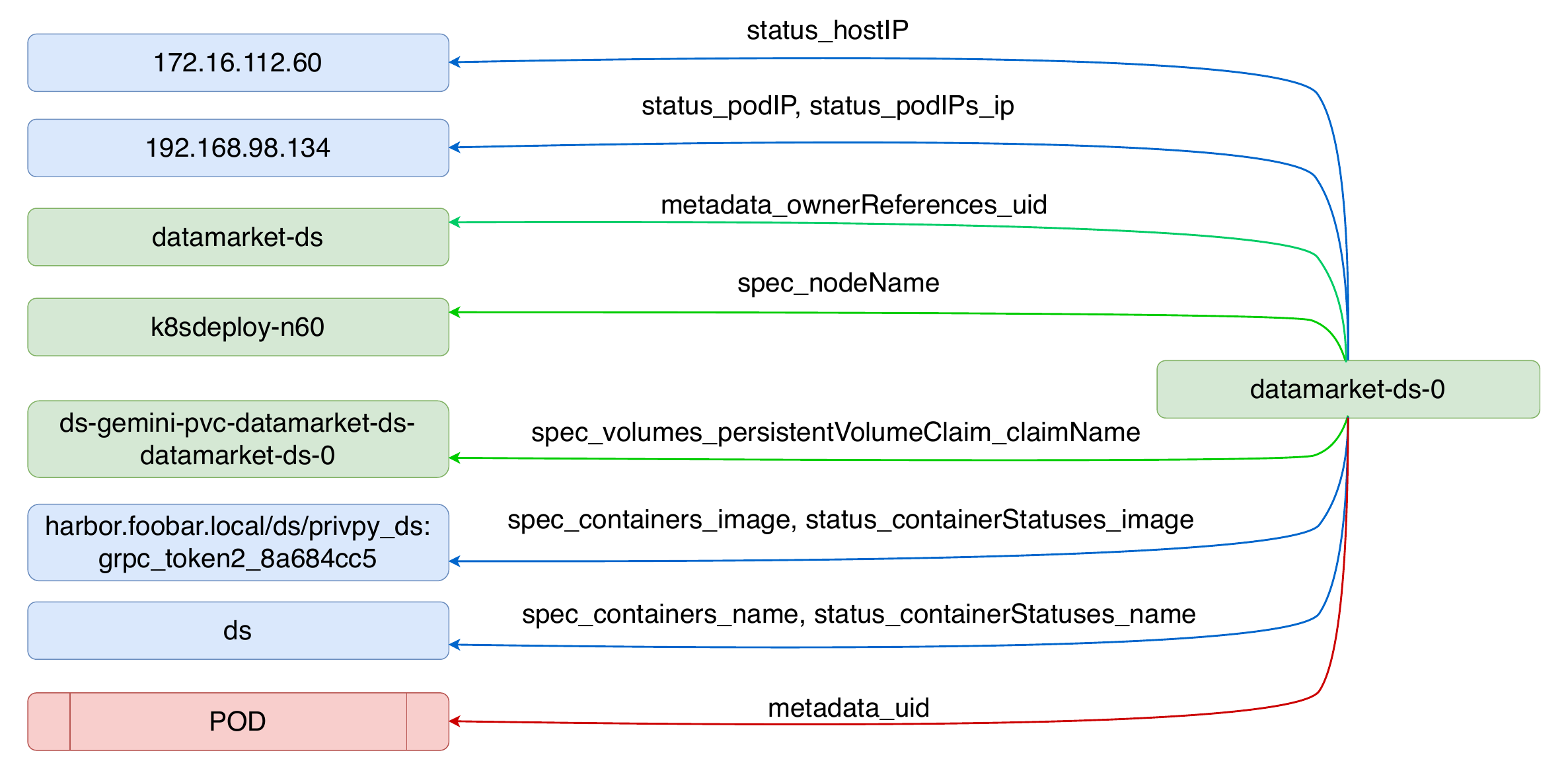}
  \caption{An example \stategraph{} snippet for a single Pod snapshot.}
  \label{fig:StateGraphFromSingleSnapshot} 
\end{figure}

The construction of the  \stategraph{}  involves a streamlined process consisting of collecting snapshots, removing duplicates, extracting and canonicalizing entities, and creating vertices and edges. An example snippet of a  \stategraph{}  is shown in  Fig.~\ref{fig:StateGraphFromSingleSnapshot}.

\begin{itemize}
 
\item{\emph{Collecting snapshots.}} We periodically collect snapshots from diverse components across multiple layers, including database (e.g., etcd in \kbs{}), compute runtime (e.g., containers in Nodes), storage systems (e.g., NFS servers and mount directories), network configurations (e.g., Calico, iptables), and images (e.g., images in Nodes and repositories).

\item{\emph{Removing duplicates.}} 
To ensure accurate tracking of state evolution, we eliminate consecutive duplicate snapshots while retaining the last one. This method allows us to trace changes, such as \texttt{Deployment} replica counts over time. Each snapshot maintains its associated time range.

\item{\emph{Extracting entities.}} 
Entities are extracted from each snapshot by identifying keys likely to represent or constitute an entity. Utilizing a rules and statistics-based method inspired by previous work ~\cite{xu2009largescale}, alongside human validation, we create a reliable reference. Keys that frequently appear with diverse values are identified as entity keys. We employ batch processing within Spark~\cite{zaharia2010spark}  to efficiently extract these key-value pairs.

\item{\emph{Canonicalizing entities.}} 
For consistent matching, entities are canonicalized. For example, transforming a \texttt{Pod}'s  reference to a \texttt{PVC} by its name into its canonical form (uid, name, namespace, kind) ensures uniformity across snapshots.

\item{\emph{Creating vertices and edges.}} 
For each snapshot, we designate a \emph{primary} entity and establish connections to other referenced entities. A vertex is created for each entity, with ReferInternal or UseExternal edges connecting the primary entity to other referenced entities. Additionally, a vertex is created for the snapshot itself, linking it to the primary entity through HasState or HasEvent edges. These edges include properties such as the snapshot’s time range and key. For a series of snapshots, we consolidate edges with identical source, destination, and key, using $t_{\text{min}}$ and $t_{\text{max}}$  to indicate the valid time range. Duplicated vertices are merged to maintain a clean and efficient graph structure.

\end{itemize}


\subsection{Extracting \metagraph{} from \stategraph{}}
\label{subsec:metagraph_design}

We construct a \emph{\metagraph{}}, similar to a class in object-oriented programming, to represent the overall structure of the \stategraph{}, which resembles an object. This \metagraph{} is directly extracted from the \stategraph{} to ensure precision, avoiding extra or missing connections that may result from using alternative sources like \kbs{} documentation (e.g., Custom Resources). Such accuracy is essential for accurately identifying instantiated statepaths for a given metapath.

To form the \metagraph{}, we create a distinct vertex for each entity kind in the \stategraph{}. For example, a \textit{Pod} vertex represents all Pods. These vertices are grouped into three categories: \emph{NativeEntity} for k8s-native entities (e.g., \texttt{Pod}, \texttt{StatefulSet}), \emph{ExternalEntity} for external entities (e.g., \texttt{container}, \texttt{image}), and \emph{Snapshot} for snapshots (e.g., \texttt{POD} for a \texttt{Pod}’s state). We retain the edge types from the \stategraph{}, categorizing them into four types: \emph{ReferInternal}, \emph{UseExternal}, \emph{HasState}, and \emph{HasEvent}. For instance, a ReferInternal edge between a \textit{Pod} vertex and a \textit{StatefulSet} vertex signifies a reference relationship.

\begin{figure}[tb]
  \centering
  \includegraphics[width=0.5\textwidth]{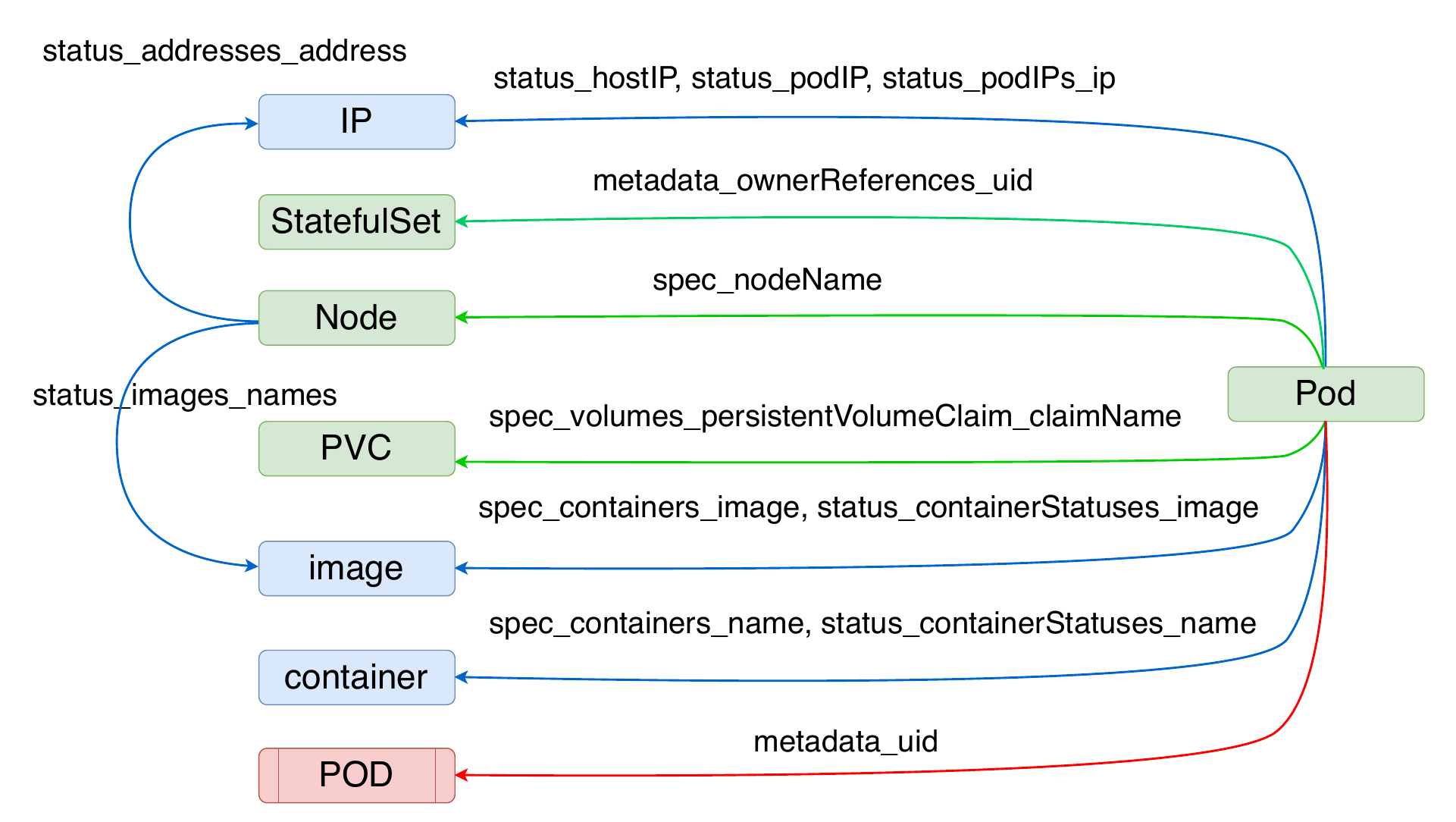}
  \caption{An Example \metagraph{} snippet encompassing the \stategraph{} snippet. }
  \label{fig:MetaGraphCoverStateGraph}
\end{figure}

The \metagraph{} construction involves the following steps: 
First, from each triplet in the \stategraph{} (comprising a source vertex, edge, and a destination vertex), we extract a quadruplet in the format (srcKind, destKind, key, type). For example, a triplet involving a \texttt{Pod} and a \texttt{StatefulSet} may produce a quadruplet like (\texttt{Pod}, \texttt{StatefulSet}, \texttt{metadata\_owner\-References\_name}, \texttt{ReferInternal}). 
Next, we remove duplicates from these quadruplets (or count their occurrences) to ensure that each unique relationship is represented once (or to collect frequency statistics). 
Then, we create vertices for each distinct srcKind and destKind. Subsequently, we establish edges between these vertices using the edge type (e.g., ReferInternal) and key (e.g., \texttt{\seqsplit{metadata\_owner\-References\_name}}) to define the relationship nature. These steps result in a \metagraph{}, as illustrated in Fig.~\ref{fig:MetaGraphCoverStateGraph}, potentially containing additional edges due to further \stategraph{} snippets from other snapshots.

\section{Evaluation}
\label{evaluation}

In this section, we take a comprehensive approach to evaluate \toolname{}, focusing on its effectiveness, module performance, and efficiency. Initially, we demonstrate \toolname{}'s capability to accurately identify root causes within \kbs{} clusters, showcasing its real-world applicability. Next, we assess the performance of each LLM module in \toolname{}, highlighting their precision and reliability. Finally, we analyze time and token costs to measure efficiency.

This comprehensive evaluation seeks to validate \toolname{}'s practical applicability, the robustness of its modules, and its operational efficiency, reaffirming its value as a powerful RCA tool in dynamic environments such as \kbs{} clusters.

\subsection{Implementation and Setup}

We develop \toolname{}  for  \kbs{}  by leveraging \gpt{} as our LLM and Neo4j~\cite{neo4j} for managing \stategraph{} and \metagraph{} databases. Initially, we build the graphs using PySpark~\cite{zaharia2010spark, pyspark} and the GraphFrames library~\cite{dave2016graphframes, graphframe} for scalability and efficiency, then seamlessly integrate them into Neo4j for robust graph queries. We implement LLM modules in \toolname{} using OpenAI's Assistants API~\cite{asst_openai, asst_azure} and executed \gpt{} on Microsoft Azure.

\toolname{} comprises approximately 5900 lines of Python code, divided into 1620 lines for data collection, 2480 lines for constructing \stategraph{} and \metagraph{} in PySpark, and 1760 lines for LLM-driven analysis using \gpt{}. Python code also generates bash commands for importing graphs into Neo4j, streamlining data integration.


For evaluation, we collect data from two production \kbs{} clusters of different sizes and versions. The first cluster, with 27 nodes running version 1.18, provides a week's worth of data. The second cluster, with 88 nodes running version 1.21, offers six months of data. After removing consecutive duplicates, we retain datasets of 13.2GB and 118.8GB, respectively, ensuring high-quality data for evaluation. Incident owners and senior \kbs{} administrators~\cite{ahmed2023recommending, zhang2024automated} are invited to formulate the ground truth and evaluate the results to ensure assessment accuracy.

\subsection{Identifying Root Cause}
\label{subsec:identify_root_cause}

We prepare test datasets comprising frequently occurring \kbs{} incidents with observable impacts, such as  \texttt{FailedCreate}, \texttt{FailedMount} and \texttt{Evicted}, which arise from different reasons. Each reason can yield various error message types; for example, \texttt{FailedCreate}  may include \texttt{Exceed\-Quota\-Job} or \texttt{Service\-Account\-Not\-Found} scenarios. Table  \ref{tab:error_message_example} provides a few examples. To ensure broad coverage, we randomly sampled these messages across different namespaces and timestamps.

\begin{table*}[htbp]
    \centering
    \scriptsize 
    \caption{Error message examples in two evaluation datasets}
    \label{tab:error_message_example}

    \begin{tabularx}{0.85\textwidth}{llX}
        \toprule
        \textbf{Reason} & \textbf{Type} & \textbf{Message} \\
        \midrule
        
Failed	&NoVolumeToMount		& Error: cannot find volume ``gen-white-list-conf'' to mount into container ``es-crontab-job'' \\ 


FailedCreate	&ExceedQuotaJob	&Error creating: pods \enquote{es-cronjob-1607637300-fpb68} is forbidden: exceeded quota: compute-resources-\baishen{}, requested: pods=1, used: pods=50, limited: pods=50 \\


FailedCreate 	&ExceedQuotaReplicaSet 	&Error creating: pods \enquote{normal-es-s31-66bf4bb56d-zp2m4} is forbidden: exceeded quota: compute-resources-\dushiting{}, requested: limits.memory=32Gi, used: limits.memory=5372Gi, limited: limits.memory=5400Gi \\


FailedCreate 	&ExceedQuotaStatefulSet 	&create Pod \yanghao{}-c1-0 in StatefulSet \yanghao{}-c1 failed error: pods \enquote{\yanghao{}-c1-0} is forbidden: exceeded quota: compute-resources-\yanghao{}, requested: pods=1, used: pods=50, limited: pods=50 \\

FailedMount	&NoSuchFileDir	&(combined from similar events): MountVolume.SetUp failed for volume  \texttt{`pvc-ca00f7a6-fb99-49a1-9881-e1f98db9297d'}: mount failed: exit status 32 Mounting command: systemd-run Mounting arguments: -{}-description=Kubernetes transient mount for  \texttt{/var/lib/kubelet/pods/ae161f5f-c87e-4c7a-aea5-51a46532ee7e/\allowbreak volumes/kubernetes.io\textasciitilde{}nfs/pvc-ca00f7a6-fb99-49a1-9881e1f98\-db9297d} -{}-scope -{}- mount -t nfs  \texttt{172.16.112.63:/mnt/k8s\_nfs\_pv/\chaimengfei{}-common\allowbreak-mysql-pvc-0-common-mysql-0-0-pvc-ca00f7a6\-fb99-49a1-9881-e1f98db9297d}  \texttt{/var/lib/kubelet/pods/ae161\-f5fc87e-4c7a-aea5-51a46532ee7e/volumes\allowbreak/kubernetes.io\textasciitilde{}nfs/pvc-ca00f7a6-fb99-49a1-9881\-e1f98db9297d} Output: Running scope as unit: \texttt{run-r5389742587e44b6db650d4ea59fb94e5.scope} mount.nfs: mounting  \texttt{172.16.112.63:/mnt/k8s\_nfs\_pv/\chaimengfei{}\-common-mysql\allowbreak-pvc-0-common-mysql-0-0-pvc-ca00f7a6-fb99-49a1\-9881-e1f98db9297d} failed, reason given by server: No such file or directory \\

FailedScheduling	&UnboundPVC		&pod has unbound immediate PersistentVolumeClaims (repeated 19 times) \\

FailedScheduling	&NodeNotAvailable	&0/87 nodes are available: 1 Too many pods, 3 node(s) had taint \{node-role.kubernetes.io/master:\}, that the pod didn’t tolerate, 38 node(s) were unschedulable, 45 node(s) didn’t match Pod’s node affinity/selector.\\

        \bottomrule
    \end{tabularx}
\end{table*}



\begin{table*}[htbp]
    \centering
     \scriptsize
     \caption{Precision of \toolname{} in two datasets}
     \label{tab:precision_two_datasets_strict}

\begin{tabularx}{0.8\textwidth}{llcccccc}
        \toprule
        \textbf{Reason} & \textbf{Type} & \multicolumn{3}{c}{\textbf{dataset-1}} & \multicolumn{3}{c}{\textbf{dataset-2}} \\
        \cmidrule(lr){3-5} \cmidrule(lr){6-8}
        & & \textbf{\#Correct} & \textbf{\#Example} & \textbf{Precision} & \textbf{\#Correct} & \textbf{\#Example} & \textbf{Precision} \\
        \midrule
  
ClaimLost	&PVLost			&- 	&- 	&-		&21	&23	&0.91 \\
Evicted	&LowOnResource	&20	&20	&1.00	&34	&35	&0.97 \\
Evicted	&NodeDiskPressure	&24	&24	&1.00	&24	&24	&1.00 \\
Failed	&AccessDenied	&39	&39	&1.00	&- 	&- 	&-\\	
Failed	&ArtifactNotFound	&20	&20	&1.00	&21	&22	&0.95 \\
Failed	&NetworkUnreachable	&21	&21	&1.00	&- 	&-	&-  \\	
Failed	&NoVolumeToMount		&18	&24	&0.75	&37	&37	&1.00 \\
FailedCreate	&ExceedQuotaJob	&37	&45	&0.82	&37	&46	&0.80 \\
FailedCreate	&ExceedQuotaReplicaSet		&18	&32	&0.56	&42	&54	&0.78 \\
FailedCreate	&ExceedQuotaStatefulSet		&19	&20	&0.95	&41	&44	&0.93 \\
FailedCreate	&ServiceAccountNotFound	&32	&32	&1.00	&40	&40	&1.00 \\
FailedMount	&ConfigMapNotFound		&43	&43	&1.00	&57	&58	&0.98 \\
FailedMount	&FailedSyncConfigMapCache	&56	&56	&1.00	&64	&64	&1.00 \\
FailedMount	&FailedSyncSecretCache		&54	&57	&0.95	&60	&60	&1.00 \\
FailedMount	&NoSuchFileDir		&47	&47	&1.00	&42	&42	&1.00 \\
FailedMount	&ObjectNotRegistered	&14	&24	&0.58	&34	&34	&1.00 \\
FailedMount	&PVCNotBound		&-	&-	&-		&39	&56	&0.70 \\
FailedMount	&SecretNotFound		&56	&56	&1.00	&55	&55	&1.00 \\
FailedMount	&ServiceAccountNotFound	&-	&-	&-	&25	&25	&1.00 \\
FailedMount	&StaleNFS			&21	&21	&1.00	&- 	&-	&-  \\		
FailedScheduling	&PVCNotFound	&-	&-	&-		&34	&34	&1.00 \\
FailedScheduling	&UnboundPVC		&10	&38	&0.26	&33	&71	&0.46 \\
OutOfpods	&NodeNotEnough		&-	&-	&-		&19	&19	&1.00 \\

        \midrule
        \textbf{Total} 	& 	& 549	 &619 & 	 	&759 &843 &  \\
        \textbf{Average} & 	& 	& 	&0.88 	& 	& 	&0.92 \\
        \bottomrule

    \end{tabularx}
\end{table*}


The precision of \toolname{} for two datasets is reported in Table~\ref{tab:precision_two_datasets_strict}. For each error message, referred to as an \emph{example}, \toolname{} attempts to generate a report up to three trials. If \toolname{} produces a report that reasonably explains the root cause based on the ground truth, it is counted as \emph{correct}. Precision is defined as the percentage of correctly explained examples. The average precision achieved is 0.88 for dataset-1 and 0.92 for dataset-2.




 
Examples such as \texttt{Evicted-LowOnResource}, \texttt{FailedCreate-ServiceAccountNotFound} and \texttt{FailedMount-NoSuchFileDir} were well explained, while \texttt{FailedCreate-ExceedQuotaReplicaSet}, \texttt{FailedMount-ObjectNotRegistered} and \texttt{FailedScheduling-UnBoundPVC} exhibited lower precision.
The primary reason is that there is a \emph{discrepancy} in data due to inconsistent snapshot.
For instance, while checking \texttt{PERSISTENTVOLUMECLAIM}, the status might show ``bound," whereas the error indicates ``Unbound." This is due to snapshots being captured every 5 minutes without ensuring consistency, potentially missing rapidly changing states. Similar issues occur in \texttt{FailedCreate-ExceedQuota*} when a pod requests CPU or memory resources. If we tolerate inconsistent snapshots, precision increases above 0.95, highlighting \toolname{}'s effectiveness using LLM.

\subsection{Effectiveness of Each LLM Module}

We closely examine the performance of the LLM-based modules in \toolname{}, with results shown in Table~\ref{tab:component_metrics_two_datasets}.


\begin{table*}[htbp]
    \centering
     \scriptsize
     \caption{Correctness metrics of each LLM module in two datasets}
     \label{tab:component_metrics_two_datasets}

\begin{tabularx}{0.75\textwidth}{lcccc}
        \toprule
        \textbf{Metric (\textit{see text})}   & \multicolumn{2}{c}{\textbf{dataset-1}} & \multicolumn{2}{c}{\textbf{dataset-2}} \\
        
        \cmidrule(lr){2-3} \cmidrule(lr){4-5}
        & \textbf{Weighted Mean} & \textbf{Arithmetic Mean} & \textbf{Weighted Mean} & \textbf{Arithmetic Mean} \\
        \midrule

\rclplain-Precision	&0.89	&0.91	&0.92	&0.95 \\
\rclplain-Precision (without knowledge) 	&0.84	&0.79	&0.91	&0.90 \\
\cqgplain-Precision	&0.95	&0.95	&0.95	&0.94 \\
\rcgplain-Conlusion-Precision		&0.92	&0.94	&0.93	&0.95 \\
\rcgplain-Command-Precision	&0.97	&0.97	&0.94	&0.94 \\
\iesplain-FPR		&0.09	&0.08	&0.10	&0.07 \\
\iesplain-FNR		&0.04	&0.04	&0.07	&0.05 \\

	\bottomrule
        
    \end{tabularx}
\end{table*}

For \rcl{}, we assess its ability to propose the most relevant destKind, rating these proposals as high, moderate, low, or unrelated. Only high-related destKinds are considered correct, and we calculate their percentage as \textit{\rclplain{}-Precision} (rows 1--2).

For \cqg{},  we show the ability to generate runnable Cypher query, we take the percentage of successfully completed queries with expected output as \textit{\cqgplain{}-Precision} (row 3).

For \rcg{}, we invite \kbs{} experts to evaluate the deduction quality from the diagnostic summary to conclusion. Deductions aligning with expert reasoning and leading to valid conclusions are deemed correct, calculated as \textit{\rcgplain{}-Conclusion-Precision} (row 4). Experts also assess the usefulness of recommended commands, with adherence to common remediation practices deemed helpful, tallied as \textit{\rcgplain{}-Command-Precision} (row 5).

For  \ies{}, we evaluate further investigation suggestions against expert labels. A false positive occurs when \ies{} labels a trial as True, indicating no further investigation is necessary, but the expert labels it as False, implying otherwise (row 6). Conversely, a false negative arises when \ies{} advises further investigation by labeling the trial as False, although the expert believes it True, requiring no additional analysis (row 7). 
False positives can hinder error message interpretation, while false negatives result in unnecessary use of computational resources.
We consider both error rates to assess \ies{}'s effectiveness.

Due to the verbose nature of \ds{} summaries, which makes semantic quality difficult to quantify, we omit this module and consider the overall precision (e.g. 0.88) as a lower bound for its performance.

We calculate both the weighted mean and arithmetic mean for each metric. Essentially, each metric has a table similar to Table~\ref{tab:precision_two_datasets_strict}. The weighted mean is computed as Total(\#Correct)/Total(\#Example) (e.g., 549/619), while the arithmetic mean is the average of the precision values (e.g., 0.88).


Table~\ref{tab:component_metrics_two_datasets} shows 
the \textit{mean \rclplain{}-Precision} ranges from 0.89 to 0.95 (row 1), indicating effective prediction of high-related destKinds. Without graph and expert guidance (i.e., Fig.~\ref{fig:graph_expert_knowledge}), this precision drops to 0.79 to 0.91 (row 2), where determining the correct volume type (e.g., \texttt{ConfigMap} or \texttt{Secret})  or case (e.g., \texttt{nfs} or \texttt{NFS}) becomes challenging. 
The \textit{mean \cqgplain{}-Precision}  of approximately 0.95 (row 3) indicates a low failure rate of 5\%, demonstrating its robust ability to accurately convert metapaths into executable queries.
The \textit{mean \rcgplain{}-Conclusion-Precision} ranges from 0.92 to 0.95 (row 4), demonstrating its ability to draw reliable conclusions, and \textit{mean \rcgplain{}-Command-Precision}  spans from 0.94 to 0.97 (row 5), indicating its effectiveness in recommending useful remediation commands. 
The \textit{mean \iesplain{}-FPR} ranges from 0.07 to 0.10 (row 6), suggesting possible oversight of cases needing retries due to occasional acceptance of a conclusion despite discrepancies. 
The \textit{mean \iesplain{}-FNR}, from 0.04 to 0.07 (row 7), shows it rarely rejects correct reports, thus minimizing unnecessary computational resource usage.

\subsection{Time and Token Cost}
\label{subsec:efficiency}

Given an error message, \toolname{} runs a max of three trials to explain the root cause, we call each trial as an \emph{attempt}.  We show the time and token costs of each attempt in Table~\ref{tab:time_token_cost_dataset_2020} and Table~\ref{tab:time_token_cost_dataset_2023} for dataset-1 and dataset-2, respectively.  

\begin{table*}[htbp]
    \centering
    \scriptsize
    \caption{Average time and token cost for each attempt of an error message in dataset-1}
    \label{tab:time_token_cost_dataset_2020}

    \begin{tabularx}{0.75\textwidth}{llcccc}
        \toprule
        \textbf{Reason} & \textbf{Type} & \textbf{TimeCost (sec)} & \textbf{PromptToken} & \textbf{CompletionToken} & \textbf{TotalToken} \\
        \midrule
	Evicted	&LowOnResource	&93.99	&119179.05	&1052.60	&120231.65 \\
	Evicted	&NodeDiskPressure	&75.14	&122300.46	&919.25	&123219.71 \\
	Failed	&AccessDenied	&110.46	&136474.10	&1486.08	&137960.18 \\
	Failed	&ArtifactNotFound	&741.95	&1335117.00	&5178.65	&1340295.65 \\
	Failed	&NetworkUnreachable	&152.99	&114172.90	&1930.10	&116103.00 \\
	Failed	&NoVolumeToMount		&115.39	&106318.08	&1347.16	&107665.24 \\
	FailedCreate	&ExceedQuotaJob	&101.72	&112012.95	&1048.98	&113061.93 \\
	FailedCreate	&ExceedQuotaReplicaSet		&84.34	&44067.00	&945.89	&45012.89 \\
	FailedCreate	&ExceedQuotaStatefulSet		&103.97	&95695.17	&1222.83	&96918.00 \\
	FailedCreate	&ServiceAccountNotFound	&79.09	&50227.56	&903.00	&51130.56 \\
	FailedMount	&ConfigMapNotFound	&46.86	&27001.02	&523.33	&27524.35 \\
	FailedMount	&FailedSyncConfigMapCache	&42.59	&19088.24	&523.38	&19611.62 \\
	FailedMount	&FailedSyncSecretCache		&85.98	&49897.12	&736.77	&50633.89 \\
	FailedMount	&NoSuchFileDir	&147.71	&189353.10	&2506.44	&191859.54 \\
	FailedMount	&ObjectNotRegistered	&57.87	&45831.43	&524.82	&46356.25 \\
	FailedMount	&SecretNotFound	&42.49	&35634.36	&563.16	&36197.52 \\
	FailedMount	&StaleNFS	&193.21	&137606.00	&2174.62	&139780.62 \\
	FailedScheduling	&UnboundPVC	&82.26	&128558.25	&994.53	&129552.78 \\
        
         \midrule	
          \textbf{Average} & 	&131.00	&159362.99	&1365.64	&160728.63 \\
           
        \bottomrule
    \end{tabularx}
\end{table*}

\begin{table*}[htbp]
    \centering
    \scriptsize
     \caption{Average time and token cost for each attempt of an error message in dataset-2}
     \label{tab:time_token_cost_dataset_2023}
    
    \begin{tabularx}{0.75\textwidth}{llcccc}
        \toprule
        \textbf{Reason} & \textbf{Type} & \textbf{TimeCost (sec)} & \textbf{PromptToken} & \textbf{CompletionToken} & \textbf{TotalToken} \\
        \midrule
        ClaimLost	&PVLost			&122.18	&46828.45	&1024.45	&47852.90 \\
	Evicted	&LowOnResource	&107.53	&61658.14	&770.92	&62429.05 \\
	Evicted	&NodeDiskPressure		&96.60	&44501.46	&668.75	&45170.21 \\
	Failed	&ArtifactNotFound		&149.82	&103740.09	&1272.74	&105012.83 \\
	Failed	&NoVolumeToMount		&147.08	&60986.00	&938.86	&61924.86 \\
	FailedCreate	&ExceedQuotaJob	&99.52	&54945.42	&816.05	&55761.47 \\
	FailedCreate	&ExceedQuotaReplicaSet		&131.67	&165871.66	&1104.28	&166975.95 \\
	FailedCreate	&ExceedQuotaStatefulSet		&118.75	&83234.30	&1266.96	&84501.26 \\
	FailedCreate	&ServiceAccountNotFound	&156.96	&145011.48	&975.20	&145986.68 \\
	FailedMount	&ConfigMapNotFound	&90.06	&27608.33	&584.45	&28192.78 \\
	FailedMount	&FailedSyncConfigMapCache	&108.28	&44007.63	&591.33	&44598.96 \\
	FailedMount	&FailedSyncSecretCache	&129.67	&54923.32	&851.10	&55774.41 \\
	FailedMount	&NoSuchFileDir	&130.71	&78662.62	&1597.67		&80260.29 \\
	FailedMount	&ObjectNotRegistered	&105.05	&33504.15	&637.12	&34141.26 \\
	FailedMount	&PVCNotBound	&119.72	&99246.44	&1135.55	&100381.98 \\
	FailedMount	&SecretNotFound	&108.35	&24171.76	&591.29	&24763.05 \\
	FailedMount	&ServiceAccountNotFound	&126.57	&31199.46	&882.65	&32082.12 \\
	FailedScheduling	&PVCNotFound	&88.83	&31818.56	&554.21	&32372.76 \\
	FailedScheduling	&UnboundPVC		&146.46	&186477.81	&1036.35	&187514.15 \\
	OutOfpods	&NodeNotEnough	&89.69	&65294.05	&718.42	&66012.47 \\
	
         \midrule	
           \textbf{Average} & 	&118.67	&72184.56	&900.92	&73085.47 \\
           
        \bottomrule
    \end{tabularx}
\end{table*}

The average time cost across attempts is approximately 2 minutes. The average total token cost varies from 73K to 161K tokens, translating to roughly \$0.19 to \$0.41 based on \gpt{} pricing~\cite{gptpricing}. Importantly, about 99\% of the total token cost comes from input prompt tokens. This high percentage is due to two main factors: (1) \ds{} processes the content of each STATE vertex, which constitutes the majority of the prompt tokens, as seen in scenarios like \texttt{Failed\-Create-Exceed\-Quota\-ReplicaSet} where it examines \texttt{REPLICASET},  \texttt{NAMESPACE}  and  \texttt{RESOURCEQUOTA}, each containing spec and status fields. (2) The output of each module is significantly shorter than the input prompt. For example, \rcl{} and \cqg{} use a prompt of approximately 50 lines (604 tokens) but yield a 3-line JSON output (41 tokens) or a 15-line Cypher query (288 tokens).

We also observe that the \texttt{Failed-Artifact\-Not\-Found} case in dataset-1 has a longer average running time due to the necessity of exhaustively verifying approximately 10.30 metapaths, whereas others typically only verify around 1.32 metapaths on average.

\section{Discussion}
\label{discuss}

We have demonstrated the effectiveness and efficiency of \toolname{}. However, capturing consistent snapshots of fast-changing resources such as CPU and memory remains challenging, highlighting a limitation of the \stategraph{}. This challenge arises from inherent difficulties in data collection, despite our efforts to estimate polling frequency. Implementing event-driven methods like triggers or logs could impose significant burdens on the target system (i.e., \kbs{}), potentially disrupting normal cluster operations.

While the state reconciliation principle is broadly applicable, some error messages necessitate more advanced checking strategies beyond merely verifying the STATE vertex's existence and content. For instance, addressing the \texttt{Failed\-Scheduling-Node\-Not\-Available}  issue, as seen in Table~\ref{tab:error_message_example}, requires aggregating data across all Nodes, even if \rcl{} predicts the destKind as \texttt{Node}. Although foundational elements are in place, further enhancements are necessary to address more complex cases. We see the use of an agent~\cite{wang2023rcagent} with tool-augmentation generation (TAG) as a promising approach to support diverse checking strategies and handle a wider range of scenarios, which we plan to explore in future work.

\section{Conclusion}
\label{conclude}
In this paper, we present \toolname{}, an innovative tool utilizing cutting-edge LLMs such as \gpt{} for root cause analysis (RCA) in \kbs{} environments. \toolname{} employs retrieval-augmented generation (RAG) from graph databases and expert prompts to boost LLM effectiveness, bypassing the need for expensive fine-tuning or extensive in-context learning demonstrations. 
Through a comprehensive evaluation on two real-world datasets from production \kbs{} clusters, we demonstrate that \toolname{} can accurately identify root causes, achieving average precision of 0.88 and 0.92, and rapidly pinpointing issues in approximately 2 minutes on average. Our findings reveal that \toolname{} identifies numerous root causes, including novel ones, thereby significantly advancing the state-of-the-art in RCA. This research highlights the potential of combining LLMs with graph databases to enhance reliability and effectiveness in dynamic cloud environments, as well as AIOps more broadly.



\bibliographystyle{IEEEtran}
\bibliography{reference}

\end{document}